\documentclass[aps,12pt,nofootinbib]{revtex4}

\usepackage{epsfig}
\usepackage{graphicx,graphics}
\usepackage{amsmath}
\usepackage{amsfonts}
\usepackage{amssymb}

\begin{document}

\title{Mixed-state  fidelity susceptibility through iterated commutator series expansion}

\author{ N. S. Tonchev}

\affiliation{Institute of Solid State Physics, Bulgarian Academy of Sciences,
1784 Sofia, Bulgaria}

\begin{abstract}

 We present  a perturbative approach to
 the problem of computation of  mixed-state fidelity susceptibility (MFS) for thermal states. The mathematical techniques  used provides an analytical expression for the MFS  as a formal expansion in terms of the thermodynamic mean values of successively higher commutators of the Hamiltonian with the operator involved  through the control parameter. That expression is naturally divided  into two parts:  the usual isothermal susceptibility  and a constituent in the form of an infinite series of thermodynamic mean values which encodes the  noncommutativity in the problem.
 If the symmetry properties of the Hamiltonian are given in terms of the generators of some (finite dimensional) algebra, the obtained expansion may be evaluated in a closed form.  This issue is tested on several popular  models, for which it is shown that the  calculations are much simpler if they are based on the properties from the representation theory of the
Heisenberg  or SU(1, 1)  Lie algebra.

\end{abstract}

\maketitle

%\noindent \emph{Keywords}: phase transitions, fidelity, fidelity susceptibility

%Uncomment for PACS numbers title message
%\pacs{ }

\section{Introduction}

The problem of similarity between quantum states (ground or thermal)  has a variety of links  to  various areas of application in quantum mechanics, statistical physics and quantum information theory \cite{NCh00,BZ06,P08, M09}. In particular, it is related to the geometrical structure of the set of the mixed quantum states under consideration. Initiated by ideas from the linear response theory and the differential-geometric approach, the concept of mixed-state fidelity susceptibility (MFS) \cite{YLG07,ZGC07} plays a prominent role in this field.

The MFS can be expressed as  the leading term in the expansion of the Uhlmann -- Jozsa fidelity \cite {U76,J94},
\begin{equation}
{\cal F}(\rho_1,\rho_2) = \mathrm{Tr}\sqrt{\rho_1^{1/2} \rho_2 \rho_1^{1/2}}, \label{defFidel}
\end{equation}
 in the case of two infinitesimally close quantum states, see, e.g., \cite{Gu10,V10}:
\begin{equation}
\chi_{F}(\rho(0)):=\lim _{h \rightarrow 0}\frac{-2\ln {\cal F}(\rho(0),\rho(h))}{h^{2}}=
-\left. \frac{\partial^{2}{\cal F}(\rho(0),\rho(h))}{\partial h^{2}}\right|_{h=0}  .
\label{dchi}
\end{equation}
 In this definition
\begin{equation}
\rho(h) = [Z_N(h)]^{-1}\exp[-\beta {\mathcal H}(h)], \label{roh}
\end{equation}
is a one-parameter family of density matrices defined for $N$-particle Hamiltonians of the form
\begin{equation}
{\mathcal H}(h)= T - h S,
\label{ham}
\end{equation}
where the Hermitian operators $T$ and $S$ do not commute in the general case,  and $Z_N(h)= \mathrm{Tr}\exp[-\beta H(h)]$ is the corresponding partition function.
Here, $h$ is a real control parameter which discriminates the termal states, i.e. the parameter in the Hamiltonian with respect to which the MFS is computed. In (\ref{dchi}), for the sake of simplicity, the reference point is taken at $h=0$.

The concept of the MFS, allows one to convey a definite geometrical  meaning to the problem, due to the fact that the  quantity:
\begin{equation}
D_{B}(\rho_1,\rho_2)=\sqrt{2-2{\cal F}(\rho_1,\rho_2)}
\label{Bures},
\end{equation}
 known as Bures distance \cite{B69} between two density matrices $\rho_1=\rho(h_1)$ and $\rho_2=\rho(h_2)$, naturally  appears as a proper geometric structure among other similarity measures, e.g. the trace distance $D_{Tr}$, or the  Hilbert-Schmidt distance $D_{HS}$ \cite{SZ03}.
If we consider the distance between two quantum states differing by infinitesimal changes  in the values of several parameters, we come to the notion of a metric tensor, i.e. the set of the coefficients of the linear element $ds^2_B$ when written  as a quadratic form in the differentials of these parameters.
For example, when a single parameter $h$ is considered, from Eqs.  (\ref{dchi}) and (\ref{Bures})
one obtains the following relation between the Bures distance and the MFS defined for two infinitesimally close states:
\begin{equation}
d^{2}_{B}(\rho(0),\rho(h))=\chi_{F}(\rho(0))h^{2} +O(h^{4}),\qquad h \rightarrow 0.
\label{infBures}
\end{equation}
 The relationship (\ref{infBures}) explains why the terms Bures metric and MFS are used  interchangeably  in the literature (see, e.g., \cite{ZVG07,AASC10,BT12,BT13}).

Further, the MFS appears  in various other contexts under different names \cite{BC94,SZ03,T96,ZVG07,Gu10}.
In particular, it coincides (apart from a numerical factor), with the
quantum Fisher information, which plays an important role in quantum metrology \cite{P2009,GA14,LXSW14}.
The recent studies of the MFS show its  importance  in statistical mechanics, quantum phase transitions and condensed-matter physics, see \cite{P2009,Gu10,V10,ZGC07,YLG07,ZVG07,ZQWS07,QC09,AASC10,S10,BT12,BT13,BTa13} along with a number of references therein.

   The paper is structured as follows. In Sec. II, we  present an expression for the  spectral representation  of the MFS which is convenient  to recast the computational problems in terms of other relevant thermodynamic quantities.
In Sec. III, by using this spectral representation a new  series expansion  for the MFS in terms  of  the so-called "Bogoljubov-Duhamel inner product of order $n$" is introduced and analyzed.  In Sec. IV, the series expansion proposed above is checked against the explicit expressions for different models. Models of this type  appear in the description of various physical systems of interest such as non-linear optics, Lipkin-Meshkov-Glick (LMG) model in the Holstein-Primakoff single boson representation and others. A summary and  discussion are given in Sec. V.

\section{Spectral representation of the  MFS }

Hereafter, for simplicity of notation, we shall write $\chi_F(\rho)$ instead of $\chi_F(\rho(0))$ and $\rho$ instead of $\rho(0)$.
 To avoid confusion, we warn the reader that  MFS
$\chi_F(\rho)$     differs from the MFS  $\chi^G_F(\rho)$  derived in \cite{AASC10} by extending the ground-state Green's function representation  to nonzero temperatures, although  in the pure states both definitions coincide. This fact has been pointed out in \cite{S10}, see also the discussion in \cite{BT12} and \cite{BT13,BTa13}.

If $T$ and $S$ commute we have (see, e.g. \cite{Gu10,V10}) the remarkable   relation between MFS $\chi_F^{cc}(\rho)$ and usual isothermal susceptibility $\chi_{(h=0)}^{cc}(\rho)$ (the superscript cc stands for commutative case)
\begin{equation}
\chi_F^{cc}(\rho)
= \frac{\beta^2}{4}\chi_{(h=0)}^{cc},
\label{Fmoda3}
\end{equation}
which establishes a relation between a theoretical information issue and a well known thermodynamic quantity.

The main computational obstacles in obtaining $\chi_{F}(\rho)$ arise in the non commutative case of the Hamiltonian (\ref{ham}).
To proceed with the calculations in the  case when the operators $T$ and $S$ do not commute, we make use of the convenient spectral representation introduced in \cite{BT12}. We assume that the Hermitian operator $T$ has a complete
orthonormal set of eigenvectors $|n\rangle$, $T|n\rangle = T_n|n\rangle$, where $n=1,2,\dots $,
with a non-degenerate spectrum $\{T_n\}$. In this basis the zero-field density matrix $\rho$ is diagonal too:
\begin{equation}
\langle m|\rho|n\rangle = \rho_n \delta_{m,n},\quad \rho_n :=e^{-\beta T_n}/Z_N(0)  ,\quad m,n =1,2,\dots .
\end{equation}
Under the above conditions, the following spectral representation for MFS was obtained \cite{BT12,BT13}:
\begin{equation}
\chi_F(\rho)
=\frac{1}{4}\beta^2 \left\{\langle (\delta S^d)^2\rangle_0 +
\frac{1}{2}\sum_{m,n, m\not=n} |\langle n|S|m \rangle|^2\frac{\rho_n
-\rho_m}{X_{mn}} \frac{\tanh X_{mn}}{X_{mn}}\right\}.
\label{FiSus2}
\end{equation}
Here $X_{mn} \equiv  \beta (T_m -T_n)/2$, and the symbol
$$\langle \cdots \rangle_0:= [Z_N(0)]^{-1}\mathrm{Tr}\{\exp[-\beta {\mathcal H}(0)]\cdots\}$$
denotes the Gibbs average value at
$h=0$, $\delta S^d = S^d - \langle S^d\rangle_0$, where $S^d$ is the diagonal part of the operator $S$,
so that
\begin{equation}
\langle (\delta S^d)^2 \rangle_0 :=\sum_m \rho_{m}\langle m|S|m \rangle^2 - \langle S\rangle_0^2.
\label{Sd}
\end{equation}

Equivalent matrix representation of the MFS can be read off from the corresponding expressions obtained in \cite{SZ03,ZVG07,AASC10} by using the identity
\begin{equation}
\tanh X_{mn}=\frac{\rho_n-\rho_m}{\rho_n+\rho_m}
\label{oid}
\end{equation}
in Eq. (\ref{FiSus2}). Representation (\ref{FiSus2}) is the starting point for the derivation of inequalities involving macroscopic quantities, like susceptibilities and thermal average values of some operator constructions \cite{BT12,BT13,BTa13}. Note that the first term in the right-hand side of eq. (\ref{FiSus2})
\begin{equation}
\chi_F^{cl}(\rho):=\frac{\beta^2}{4} \langle (\delta S^d)^2\rangle_0,
\label{FScl}
\end{equation}
describes the classical contribution (known also as Fisher-Rao term \cite{ZVG07}, cf. with (\ref{Fmoda3})) to the MFS, while the second term
\begin{equation}
\chi_F^{q}(\rho):=\frac{1}{4}\beta^2 \left[\frac{1}{2}\sum_{m,n, m\not=n} |\langle n|S|m \rangle|^2\frac{\rho_n
-\rho_m}{X_{mn}} \frac{\tanh X_{mn}}{X_{mn}}\right]
\label{FiSus2a}
\end{equation}
represents the  quantum  contribution which vanishes when the operators $T$ and $S$ commute.

In what follows we shall advocate that in lieu of (\ref{FScl}) and (\ref{FiSus2a})  it is in some sense most natural to divide the MFS in the following two parts:  the usual (quantum)isothermal susceptibility  and a part which represents an infinite series of thermodynamic mean values encoding  the effect of the  noncommutativity in the problem. In other words our aim is to obtain the quantum counterpart of the relation (\ref{Fmoda3}).

\section{Series representation of the MFS}

  In quantum physics, over the years, an approach to problems with noncommuting operators was used, which goes back to Feynman's "disentangling". Essentially, it consists  in to working with expansions in terms of successively higher commutators of the operators involved. Some time this procedure,  called "expansion in iterated commutators",  gives  results in a very compact notation \cite{V68}. The method has been developed  in different directions by many authors (for a review see \cite{P07}).    In our case the iterated commutators are:
\begin{equation}
R_{0}\equiv R_{0}(S)= S,\quad  R_{1}\equiv R_{1}(S) =[T,S],\;
\dots, \; R_{n} \equiv R_{n}(S)= [T,R_{n-1}(S)].
\label{R}
\end{equation}

 The aim of calculations presented in  this section is to convert into equivalent form  the MFS (\ref{FiSus2}) as
an expansion in iterated commutators. This can be achieved by using the functionals $F_{2n}(S;S)$, defined  earlier in \cite{BT11} by their spectral representation
\begin{equation}
F_{2n}(S;S) \equiv 2^{2n-1}[Z(0)]^{-1} \sum_{ml}|\langle m|S|l \rangle |^{2}|e^{-\beta T_{l}}
- e^{-\beta T_{m}}||X_{ml}|^{2n-1}.
\label{BT10a}
\end{equation}
Note that $F_{0}(S;S)$ coincides with the Bogoliubov-Duhamel inner product(see e.g. \cite{BT11,DLS}):
\begin{eqnarray}
F_{0}(S;S) &:=& [Z(0)]^{-1}\sum_{m ,l}{}^\prime |\langle m|S|l\rangle|^2
\frac{e^{-\beta T_{m}} - e^{-\beta T_{l}}}
{\beta[T_{l}-T_{m}]}\nonumber \\ &+& [Z(0)]^{-1}\sum_{l}{\mathrm e}^{-\beta T_{l}}
\langle l|S|l\rangle|^2,
\label{M51}
\end{eqnarray}
where the prime in the double sum means that the term with $l=m$ is excluded. Recall  that due to the relations
\begin{equation}
\chi_{(h=0)}=\frac{1}{\beta^2}\frac{\partial^2 \ln Z(h)}{\partial^2 h}|_{h=0} = F_{0}(\delta S;\delta S),
\label{fed}
\end{equation}
$F_0( \delta S;\delta S)$ is exactly the isothermal susceptibility $\chi_{(h=0)}$, see, e.g., \cite{DLS,BT11}.
Here the notation  $\delta S \equiv S -\langle S\rangle_0$ is used.

In a basis independent form  one has
\begin{equation}
F_{2n}(S;S):=
\beta^{2n-1}\langle[R_{n}^+,R_{n-1}]\rangle_{0}, \quad n=0,1,2,3,\dots,
\label{FRR}
\end{equation}
where $R_n^+$ denotes the Hermitian conjugate of $R_n$ and by definition
$R_{-1}\equiv X_{ST}$ is a solution of the operator equation
\begin{equation}
 S = [X_{ST},T].
 \label{oe}
\end{equation}

 It is easy to see that if $T=T^+$ and $S=S^+$ (as it is in our case), we have $R_{n}^+=(-1)^nR_{n},\quad
 n=0,1,2,...$ and so
equation (\ref{FRR}) can be recast, due to the cyclic property of the trace operation,  in the equivalent form
\begin{equation}
F_{2n}(S;S)=2(-1)^n\beta^{2n-1}\langle R_{n}R_{n-1}\rangle_{0}=2(-1)^{2n+1}\beta^{2n-1}\langle R_{2n-1}R_{0}\rangle_{0},\quad n=0,1,2,\dots
\label{50p}
\end{equation}
In some cases Eq. (\ref{50p}) is more convenient than Eq. (\ref{FRR}).
In the remainder, in view of relations (\ref{FRR}), the functional $F_{2n}(S;S)$ will be called "Bogoljubov-Duhamel inner product of order $n$".

By inserting the series expansion
\begin{equation}
\frac{\tanh x}{x}
=  1-\frac{1}{3}x^2+\frac{2}{15}x^4-\frac{17}{315}x^6 + ...
= \sum _{n=0}^{\infty}a_n x^{2n} ,\qquad |x|<\frac{\pi}{2},
\label{tanh}
\end{equation}
where
\begin{equation}
a_n = \frac{2^{2n+2}(2^{2n+2}-1)}{(2n+2)!}\; B_{2n+2}
\end{equation}
and $B_{2n}$ are the Bernoulli numbers,
into expression (\ref{FiSus2}) for the MFS, we obtain our basic  formula
\begin{eqnarray}
\chi_F(\rho) &=& \beta^2\sum_{n=0}^{\infty }\frac{a_n}{2^{2n+2}}F_{2n}(\delta S;\delta S)
\nonumber \\  &=& \frac{\beta^2}{4}\left\{F_0(\delta S; \delta S)+
\sum_{n=1}^{\infty }\frac{a_n}{2^{2n}}F_{2n}(S;S)\right\}.
\label{druga}
\end{eqnarray}
The first term in the rhs of (\ref{druga}) is obtained using the relation
%Now, from (\ref{Sd}) and (\ref{M51}) one  immediately obtains
\begin{eqnarray}
F_0(\delta S; \delta S) &=& F_0(S;S)-
|\langle S\rangle_0|^2 \nonumber \\  &=&
\frac{1}{2}[Z(0)]^{-1}
\sum_{m,l}{}^\prime |\langle m|S|l\rangle|^2\frac{{\mathrm e}^{-\beta T_{m}} - {\mathrm e}^{-\beta T_{l}}}{X_{lm}} + \langle (\delta S^d)^2 \rangle_{0}.
\label{Fluc2}
\end{eqnarray}
which follows from Eqs. (\ref{Sd}) and (\ref{M51}).
In the second term we have used that the terms containing diagonal matrix elements of the operator
$S$, i.e., with $m=l$ in (\ref{BT10a}), vanish in all $F_{2n}$ with $n\geq 1$.

Let us note that if one takes into account only the first term  in the rhs of (\ref{druga}) one can show with the aid of Eq.  (\ref{Fluc2})) that the result coincides with that obtained in Ref. \cite {QC09} by estimation based on the Trotter-Suzuki formula.
However, it has been pointed out that this approximation might not be valid at low temperatures (see also
formula (200) in \cite{Gu10} and the comment therein).

Clearly, the series representation (\ref{druga}) of $\chi_F(\rho)$ is correctly derived
provided the condition $\beta|T_l -T_m| < \pi$ for absolute convergence of the series (\ref{tanh}) with
$x = \beta|T_l -T_m|/2$ holds. This condition could be satisfied for models with a bounded spectrum
of $T$ and small enough $\beta$. However, the formal series (\ref{druga}) may happen to
be absolutely convergent by itself, even for models with unbounded from above spectrum $\{T_m,\;
m=1,2,3,...\}$ which violates the condition $\beta|T_l -T_m| < \pi$. We conjecture that in such cases
(\ref{druga}) yields a proper definition of the MFS $\chi_F(\rho)$. This issue
will be further examined in the next section by the examples of several popular models.

\section{Test by special  models}

  Here we shall demonstrate that if the Hamiltonian can be presented  as a set of Lie algebra elements the underlaying symmetry of the Hamiltonian  may be efficiently explored in order to obtain  a closed expression for the MFS. A similar  idea was already provided in \cite{T11} where the authors used the specific Hamiltonian  representation (in the Cartan-Weyl basis) to evaluate the zero-temperature fidelity susceptibility for the Lipkin-Meshkov-Glick model, the two-dimensional XXZ model and the Bose-Einstein condensate model.
We shall consider a family of Hamiltonians expressed in terms of the generators  of a polynomial deformation of the Heisenberg and $SU(1,1)$ Lie
algebra which are employed in various physical problems (for definitions  and a partial list of references, see \cite{KC09,LYZ10,LLZ11,Z13}). In this case, after a proper choice of the control  parameter,   the Bogoljubov-Duhamel inner product of order $n$  can be obtained order by order, and, in practice, the infinite summation in (\ref{druga}) may become very simple to perform.

 Following \cite{LYZ10,LLZ11,Z13}, we consider  the class of polynomial
algebras of degree $k-1$ defined by the commutation relations
\begin{equation}
[Q^0,Q^{\pm}]= \pm Q^{\pm},\qquad [Q^+,Q^{-}]= \Phi_k(Q^0)-\Phi_k(Q^0-1),
\label{KR1}
\end{equation}
where the structure function
\begin{equation}
\Phi_k(Q^0)=-\Pi_{i=1}^{k}\left(Q^0+\frac{i}{k}-\frac{1}{k^2}\right)
\end{equation}
is a $k^{th}$-order polynomial in $k$. We shall consider the following  Hamiltonian  \cite{Z13}:
\begin{equation}
{\mathcal H}(h)=k\omega\left(Q^0-\frac{1}{k^2}\right)+h\sqrt{k^k}(Q^+ + Q^-),\quad k=1,2,...,
\label{HAQ}
\end{equation}
In this case we take $T={\mathcal H}(0)=k\omega\left(Q^0-\frac{1}{k^2}\right)$ and $S=\sqrt{k^k}(Q^+ + Q^-)$.
%It is evident that the classical part of
%the fidelity susceptibility (\ref{FScl}) equals zero, because the diagonal matrix elements of $S$
%vanish in the basis of the eigenvectors of $T$.
From the operator equation (\ref{oe}) and after direct commutations one readily finds
\begin{eqnarray}
R_{-1} &=& -\frac{\sqrt{k^k}}{k\omega}\left(Q^+ - Q^-\right),\quad  R_0 = \sqrt{k^k}(Q^+ + Q^-),\quad
R_1=k\omega  \sqrt{k^k} (Q^+ - Q^-),\; \dots \;\nonumber \\
R_n &=&[(k\omega)^n \sqrt{k^k}[ Q^+ + (-1)^n Q^-],\quad
R_{2n-1}= [(k\omega)^{2n-1} \sqrt{k^k}[ Q^+ - Q^-].
\label{RcomR}
\end{eqnarray}
Inserting $R_0$ and $R_{2n-1}$ in (\ref{50p}), we obtain
\begin{equation}
F_{2n}(S;S)=-2(k\beta\omega)^{2n-1}{\mathcal K}(k),\quad n=0,1,2,...,\quad k=1,2,...,
\label{K}
\end{equation}
where
\begin{equation}
{\mathcal K}(k)=k^k\langle(Q^{+} - Q^-)(Q^{+} + Q^-)\rangle,\qquad k=1,2,...,
\label{MK}
\end{equation}
and here and below $\langle \dots \rangle$ denotes a thermal-equilibrium average with Hamiltonian ${\mathcal H}(0)$.

Now, by inserting the result (\ref{K}) into the series (\ref{druga}), we obtains
\begin{eqnarray}
\chi_F(\rho)& = &\frac{\beta^2}{4}F_0(\delta S;\delta S)-
2\beta^2{\mathcal K}(k)\sum_{n=1}^{\infty}\frac{a_n}{2^{2n+2}}(k\beta \omega)^{2n-1}\nonumber \\
&=& \frac{\beta^2}{4}\left\{\chi_{(h=0)} -\frac{{\mathcal K}(k)}{(k\beta \omega/2)}
\left[\frac{\tanh(k\beta \omega/2)}{(k\beta \omega/2)}- 1\right]\right\}, \quad
0< \beta \omega < \frac{\pi}{k}.
\label{Fmod1}
\end{eqnarray}

For further applications it is convenient to present (\ref{Fmod1}) in an alternative form. Applying (\ref{K}) for $n=0$  and using definition (\ref{Fluc2}), one finds after the cancelation in (\ref{Fmod1}),
that for every $k=1,2,...$,
\begin{eqnarray}
\chi_F(\rho)
= -\frac{\beta^2}{4}\left\{\frac{{\mathcal K}(k)}{(k\beta \omega/2)}
\left[\frac{\tanh(k\beta \omega/2)}{(k\beta \omega/2)}\right] + k^k|\langle (Q^++Q^-) \rangle|^2 \right\},\quad
0< \beta \omega < \frac{\pi}{k}.
\label{Fmod2}
\end{eqnarray}
Remarkably, the results (\ref {Fmod1}) and (\ref {Fmod2}) do not require knowledge of  the concrete realization of the algebra closed  by the operators $Q^{\pm},Q^0$.

Note that the condition $\beta \omega >0$ is necessary for the convergence of the
sums in the functionals $F_{2n}(S;S)$ and the condition $k\beta \omega < \pi$ is
required for the convergence of the infinite sum over $F_{2n}$ in the first line of
(\ref{Fmod1}). Therefore to achieve a complete solution we should  also provide  analytic  continuation of  $\chi_F(\rho)$
to the whole positive semiaxis. In what follows we shall give explicit results for the above formulas  in some special cases of well established and frequently used physical models.

\subsection{The $k^{th}$-order harmonic generation model}

It is shown in \cite{LYZ10} that the algebra defined by Eqs.(\ref{KR1})  has an infinite dimensional irreducible unitary representation given by the following one-mode boson realization:
\begin{equation}
Q^+ \equiv Q^+(k)= \frac{1}{(\sqrt{k})^k} (b^+)^k,\quad Q^-\equiv Q^-(k) = \frac{1}{(\sqrt{k})^k} b^k,\quad Q^{0}\equiv Q^0(k)=\frac{1}{k} \left(b^+b + \frac{1}{k}\right),
\label{pZ}
\end{equation}
which we shall use in our further calculations.
Thus the Hamiltonian of the model takes the form \cite{Z13}
\begin{equation}
{\mathcal H}(h)=\omega b^+b + h [(b^+)^k + b^k],\qquad \omega >0,\quad k=1,2,3,...
\label{HAk}
\end{equation}
where $b, \;b^+$ are bosonic operators obeying the canonical commutation
relations.
The k = 1 and
k = 2 cases of (\ref{HAk}) give the Hamiltonians of the displaced and single-mode squeezed
harmonic oscillators \cite{EB02,BE02}, respectively. The Hamiltonian (\ref{HAk}) for $k=2$ is also known as Lipkin-Meshkov-Glick (LMG) model in the Holstein-Primakoff single boson representation (see e.g. \cite{Gu10} and refs. therein)
and all the result obtained here can be related to this field.

 Taking into account that the Hamiltonian ${\mathcal H}(0)$ is diagonal and invariant under the gauge transformation $ b^\pm\rightarrow b^\pm {\mathrm e}^{\pm\varphi} $ and using commutation relations (\ref{KR1}) expression (\ref{MK}) transforms into  more convenient form
\begin {equation}
{\mathcal K}(k)=k^k[\langle\Phi_k(Q_0)\rangle - \langle\Phi_k(Q_0-1)\rangle],
\label{MK1}
\end{equation}
and
    $|\langle(Q^+ + Q^-)\rangle|^2 =0$.   Thus the isothermal susceptibility $\chi_{(h=0)}$ of the model is
\begin{equation}
\chi_{(h=0)}=-(k\beta \omega/2)^{-1}{\mathcal K}(k).
\label{chi}
\end{equation}
Finally,  we obtain from (\ref{Fmod2}) the result
\begin{eqnarray}
\chi_F(\rho)
= \frac{\beta^2}{4}\left\{
\left[\frac{\tanh(k\beta \omega/2)}{(k\beta \omega/2)}\right]\chi_{(h=0)}\right\},\quad 0< \beta \omega < \frac{\pi}{k},\quad k=1,2,3,...
\label{Fmod3}
\end{eqnarray}

Thus,  the relation between the MFS and the isothermal susceptibility is a renormalized version of that obtained for
the commutative case (compare with formula (\ref{Fmoda3})). The quantum features  of the model are encoded in the function $\frac{\tanh(k\beta \omega/2)}{(k\beta \omega/2)}$ in front of the
isothermal susceptibility and in the isothermal susceptibility itself which in the case is the quantum counterpart of the classical one.
Since ${\mathcal H}(0)$ is diagonal it is possible  to calculate the thermodynamic mean value ${\mathcal K}(k)$ in Eq. (\ref{chi}).
Let us consider the cases $k=1$ and $k=2$.

\subsubsection{Shifted harmonic oscillator ($k=1$)}

This is the simplest but nevertheless a didactic example.
In the case  we have for the energy eigenvalues of ${\mathcal H}(h)$
\begin{equation}
E(n)=\omega\left(n-\frac{h^2}{\omega^2}\right),\quad n=0,1,2,...
\end{equation}
For the partition function one has the sum of the geometric progression
\begin{equation}
Z(h)={\mathrm e}^{\beta\frac{h^2}{\omega}}\sum_{n=0}^{\infty}{\mathrm e}^{-\beta \omega n}=
\frac{{\mathrm e}^{\beta\frac{h^2}{\omega}}}{1-{\mathrm e}^{-\beta \omega}}.
\end{equation}

From Eq. (\ref{fed}) it immediately follows that $ F_0(\delta S;\delta S)=2/(\beta\omega)$, and  from  (\ref{MK1}) it follows that ${\mathcal K}(1)=-1$. Finally,
the result for the fidelity susceptibility of model (\ref{HAk}) at $h=0$ is
\begin{equation}
\chi_F(\rho) =\frac{\tanh(\beta \omega/2)}{\omega^2}, \qquad 0< \beta \omega < \pi.
\label{hiFA}
\end{equation}
Here the following comments are in order.
Expression (\ref{hiFA}) may also be derived if one turn backs to the original spectral representation (\ref{FiSus2})
of the fidelity susceptibility. First of all we note that $\langle (\delta S^d)^2\rangle_0 =0$,
since $S$ has no diagonal elements with respect to the eigenvectors of $T$.  In other words for this model the
classical part of the fidelity susceptibility (\ref{FScl}) equals zero. Next, only two
terms in the double sum  of the quantum part (\ref{FiSus2a}) remain nonzero: those with indices $(m,n)= (l,l-1),\, (l-1,l)$ for
which $|\langle l|S|l-1\rangle|^2 = |\langle l-1|S|l\rangle|^2 = l$, and $X_{l,l-1} = -
X_{l-1,l} =\beta \omega /2$, we are left with
\begin{equation}
\chi_F(\rho) =[Z(0)]^{-1}\frac{\tanh(\beta \omega/2)}{\omega^2}
\left({\mathrm e}^{\beta\omega} -1\right)\sum_{l=0}^{\infty}l {\mathrm e}^{-\beta\omega l}=
\frac{\tanh(\beta \omega/2)}{\omega^2}, \qquad \beta \omega >0.
\label{hiFAb}
\end{equation}
This result confirms the fact  that the  expression for the fidelity susceptibility can be
obtained by analytical continuation of the function (\ref{hiFA}) to the whole positive
semiaxis.

\subsubsection{Single-mode squeezed harmonic oscillators ($k=2)$}

In this case the operators $Q^\pm, Q^0$ form the $SU(1, 1)$  Lie algebra. As a matter of fact, the spectrum  of ${\mathcal H}(h)$  for $k=2$ can be determined explicitly \cite{Z13,EB02,BE02}. One has for the energy eigenvalues of ${\mathcal H}(h)$ the result:
 \begin{equation}
E(m)=-\frac{1}{2}\omega +\left[m + \frac{1}{2}\right]\omega\Omega,\quad m=0,1,2,...,\quad
\end{equation}
where $\Omega=\sqrt{1-\frac{4h^2}{\omega^2}}$.
For the partition function  one obtains
\begin{equation}
Z(h)=\sum_{m=0}^{\infty}{\mathrm e}^{-\beta E(m)}=
{\mathrm e}^{\frac{\beta\omega}{2}(1-\Omega)}\sum_{n=0}^{\infty}{\mathrm e}^{-\beta m\omega \Omega}=\frac{{\mathrm e}^{\frac{\beta\omega}{2}}}{2\sinh\frac{\beta\omega\Omega}{2}}.
\label{zsq}
\end{equation}
Here, to obtain the last equality in Eq.  (\ref{zsq}), one needs the stability condition $\Omega>0$.
Setting  (\ref{zsq}) in (\ref{fed}) one obtains
\begin{equation}
F_0(\delta S; \delta S)=\frac{2}{\beta \omega}\coth\frac{\beta\omega}{2}.
\label{sqF}
\end{equation}
Using (\ref{pZ}) and (\ref{MK1}) one obtains
\begin{equation}
{\mathcal K}(2)= -2\coth\frac{\beta\omega}{2}.
\label{sqK}
\end{equation}
Substituting  (\ref{sqK}) into (\ref{chi}), we obtain from (\ref{Fmod3}):
\begin{equation}
\chi_F(\rho)=\frac{1}{2\omega^2}\coth\frac{\beta\omega}{2}\tanh (\beta\omega),\quad 0<\beta\omega <\frac{\pi}{2}.
\label{k2}
\end{equation}
Indeed, by analytical continuation of the function (\ref{k2}) to the whole positive
semiaxis one can remove the conditions imposed on  $\beta $ and $\omega$.

The fidelity susceptibility (\ref{k2}) coincides (up to a factor of four) with the corresponding element of the Bures (or Statistical  Distance) metric
for squeezed thermal states obtained in \cite{T96}. This is seen by using the relation $ \coth (\beta \omega/2)=[\cosh \beta \omega +1]/\sinh \beta \omega$ in (\ref{k2}). The quite different  approach used in \cite{T96} requires a Schur factorization  in order to perform  the square root in the definition of the fidelity which  is not a trivial task even in this simple case.

\subsection{The shifted oscillator model interacting with one fermion mode}

Let us consider a  version of model (\ref{HAk}), setting for simplicity  $k=1$ and including interaction with one fermion mode.
The Hamiltonian of  the model is:
\begin{equation}
{\mathcal H}(h)=ma^+a+\omega b^+b+ ga^+a(b^+ + b) + h(b^+ + b),
\label{MB}
\end{equation}
where $\{a,a^+\}=1$ and $[b,b^+]=1$.
It described a fixed particle of energy $m$ interacting
with a charged oscillator in constant electric field.The interaction occurs only when
the state is occupied, i.e. $a^+a=1$. The many boson mode version of  the model is used for describing a large
variety  of effects in  solid state physics \cite{M 2000}.

Defining the  operators $T$ and $S$  in (\ref{ham}) as follows:  $T=ma^+a+\omega b^+b+ ga^+a(b^+ + b)$ and $S=b^+ + b$, after simple algebra we obtain $F_{2n}(\delta S;\delta S)=2(\beta \omega)^{2n-1},n=1,2,...$ (the same result as in the case of the shifted harmonic oscillator model).
Then taking into account the definition (\ref{druga}),  it is easy to see that

\begin{equation}
\chi_F(\rho)=\frac{\beta^2}{4}\left\{\chi_{h=0}+\frac{1}{(\beta\omega/2)}\left[\frac{\tanh(\beta \omega/2)}{(\beta \omega/2)} - 1\right]\right\},
\qquad \beta \omega < \pi.
\label{er1}
\end{equation}
This equation is a particular case  in form (with ${\mathcal K}(1)=-1$) to Eq.(\ref{Fmod1}). Evidently, the  further investigation of the  structure of the MFS is hampered  by the complicated first term in Eq. (\ref{er1}).  Its calculation is described in the Appendix. The result is
\begin{equation}
\chi_{h=0}=\frac{1}{(\beta \omega/2)} + \left(\frac{g}{\omega}\right)^2\frac{1}{\cosh^2[(\beta\omega/2)(m/\omega-g^2/\omega^2)]}.
\label{fso1}
\end{equation}
Finally from (\ref{er1}) and (\ref{fso1}), for the  MFS of the model under consideration we get:
\begin{equation}
\chi_F(\rho)=\frac{\beta^2}{4}\left\{ \left(\frac{g}{\omega}\right)^2\frac{1}{\cosh^2[(\beta\omega/2)(m/\omega-g^2/\omega^2)]}+
\frac{1}{(\beta\omega/2)^2}\tanh(\beta \omega/2)\right\},\qquad \beta \omega < \pi.
\label{fer1a}
\end{equation}

Here the following comments are in order. The first and the second terms in the rhs of Eq. (\ref{fer1a}) are exactly the classical and quantum part of the MFS as one can check  after some tedious calculations based on the  spectral representation (\ref{FiSus2})  of the MFS
expressed in terms of the eigenbasis of ${\mathcal H}(0)=T$. The  second term in Eq. (\ref{fer1a})  coincides with the MFS of the shifted harmonic oscillator model, see Eq. (\ref{hiFA}), and confirms  the result (\ref{fer1a}) in the wider interval of the whole positive semiaxis.

\section{Summary and discussion}

In the literature, see, e.g., \cite{ZVG07,AASC10}, it is commonly accepted to cast the MFS in a form that distinguishes the  classical and quantum contributions. Here, the announced approach suggests the use of the symmetry aspects in the computation of the MFS. The  iterated commutator expansion (\ref{druga})  is naturally divided  into two parts:  the usual isothermal susceptibility  and  a constituent which represents  the generic  noncommutativity of the problem.

The MFS is presented here as a series in terms enumerated by  the number  $n$ of iterated commutators between $T$ and $S$ in Hamiltonian (\ref{ham}).
 As a starting point in (\ref{druga})  we take  the usual isothermal susceptibility which is related to $n=0$.
The appearance of the iterated commutators (terms with $n>0$) is a reminiscence of a disentangling procedure which is a well known and useful tool in quantum mechanics, quantum field theory, optics, etc. \cite{P07}.

If the Hamiltonian is a linear form of the generators of a representation of some (finite dimensional) Lie algebra, the  obtained series expansion can be used in a rather simple  way to obtain  closed-form expressions.
Indeed, it is a consequence of the dynamical symmetry algebra of Hamiltonian (\ref{HAk}),  spanned by the operators $\{Q^{0},Q_{-},Q^{+} \} $  with the commutation relations (\ref{KR1}), that enables one to obtain in a closed form the functions $F_{2n}( S;S)$,  see Eq. (\ref{K}).
If the Hamiltonian ${\mathcal H}(h)$ can be diagonalized for $h=0$,  the values  of ${\mathcal K}(k)$ and $\langle S \rangle$ can be calculated relatively easy in virtue of this property.

In our approach one has to accomplish  two different steps : the first one is to find a representation of (\ref{druga})  in terms of
some known functions, and the second one is to perform an analytic extension in order to remove the restrictions imposed by the convergence conditions. Both  steps are directly checked in the models considered above. The corresponding expression  (\ref{druga}), obtained in the domain of validity  on perturbative expansion, under a subsequent  analytic extension   coincides precisely with the
nonperturbative  expression as one can see from the comments after the results (\ref{hiFA}) and (\ref{k2}). Consequently, the MFS understood as an analytic continuation is defined for
all values of the parameters of the Hamiltonian under consideration.

\section*{Acknowledgement} It is a pleasure to thank  J.G. Brankov for helpful discussions and  careful reading of the text.

\section*{Appendix}

Here we shall calculate the isothermal susceptibility  $F_{0}(\delta S;\delta S)\equiv\chi_{h=0}$ for the model Hamiltonian (\ref{MB}).  First, we introduce the shifted boson operators $\tilde{b}^\pm=b^\pm + h/\omega$ . Then,
instead of Hamiltonian (\ref{MB}) we have
\begin{equation}
\tilde{{\mathcal H}}(h)=(m -2g\frac{h}{\omega})a^+a+\omega \tilde{b}^+\tilde{b}+ ga^+a(\tilde{b}^+ + \tilde{b}) -\frac{h^2}{\omega} .
\label{sh}
\end{equation}
Further, following \cite{M 2000}, we use the unitary transformation $U(\lambda)=\exp[-\lambda a^+a(\tilde{b}^+ - \tilde{b})]$, where $\lambda=-\frac{g}{\omega}$, to introduce the Bose-operators
\begin{equation}
b^\pm(\lambda):=U(\lambda)\tilde{b}^\pm U^{+}(\lambda)=\tilde{b}^\pm +\lambda a^+a,
\label{b1}
\end{equation}
and the Fermi operators
\begin{equation}
a^\pm(\lambda):=U(\lambda)a^\pm U^{+}(\lambda)=a^\pm \exp[\mp\lambda(\tilde{b}^+-\tilde{b})],
\label{f12}
\end{equation}
 in term of which we have the unitary equivalent Hamiltonian
\begin{equation}
\tilde{\tilde{{\mathcal H}}}(h)=U(\lambda)\tilde{{\mathcal H}(h)}U^+(\lambda)=\left(M -2g\frac{h}{\omega}\right)a^+(\lambda)a(\lambda)+\omega b^+(\lambda)b(\lambda)+ ga(\lambda)^+a(\lambda)[b(\lambda)^+ + b(\lambda)] -\frac{h^2}{\omega}.
\label{ssh}
\end{equation}
It is easy to see that after setting  the rhs of  Eqs. (\ref{b1}) and (\ref{f12}) into Eq. (\ref{ssh}),  Hamiltonian (\ref{ssh}) is equivalent  to the following diagonal one :
\begin{equation}
\tilde{\tilde{{\mathcal H}}}(h)=U(\lambda)\tilde{{\mathcal H}(h)}U^+(\lambda)=\left(m-\frac{g^2}{\omega}-2g\frac{h}{\omega}\right)a^+a+\omega \tilde{b}^+\tilde{b} -\frac{h^2}{\omega} .
\label{df1a}
\end{equation}

In order to obtain (\ref{df1a}) the following relations have been used:
\begin{equation}
a^+(\lambda)a(\lambda)=a^+\exp[+\lambda(b^+-b)]a \exp[-\lambda(b^+-b)]=a^+a,\quad (a^+a)^2=(a^+a).
\end{equation}
Further, we  use the unitary equivalent Hamiltonian
\begin{equation}
\tilde{\tilde{{\mathcal H}}}(h)=M(h)a^+a+\omega \tilde{b}^+\tilde{b} -\frac{h^2}{\omega} , \qquad M(h):=m-\frac{g^2}{\omega}-2g\frac{h}{\omega},
\label{df1ab}
\end{equation}
instead of (\ref{MB}).
Now the fermion and boson parts of the considered Hamiltonian are completely separated. The interesting  point (well known from
\cite{M 2000, T70}) is that the operators $a^\pm$ and $b^\pm$ in $\tilde{\tilde{{\mathcal H}}}(h)$ are the same
that enter into $\tilde{{\mathcal H}}(h).$  In other words the interaction between the bosonic oscillator and the fermion mode simply renormalizes the mass of the free fermion under  the rule $m\rightarrow M(h)$.

Thus the partition function of the model is:
\begin{equation}
Z(h)=\exp\left(\beta \frac{h^2}{\omega}\right)\mathrm{Tr}\exp [-\beta M(h) a^+a]\, \mathrm{Tr}\exp[-\beta\omega \tilde{b}^+\tilde{b}].
\end{equation}
Finally
\begin{equation}
Z(h)=\exp\left(\beta \frac{h^2}{\omega}\right)\{1+\exp[-\beta M(h]\}\,[1-\exp(-\beta \omega)]^{-1}
\end{equation}
and the free energy is
\begin{equation}
f(h)=-\beta^{-1}\ln Z(h)=-\frac{h^2}{\omega}-\beta^{-1}\ln\{1+\exp[-\beta M(h]\}+\beta^{-1}\ln[1-\exp(-\beta \omega)].
\label{fe2}
\end{equation}
By definition
\begin{equation}
F_{0}(\delta S;\delta S):=-\frac{1}{\beta}\frac{\partial^2 f(h)}{\partial^2 h}|_{h=0}=\frac{2}{\beta \omega} + \left(\frac{g}{\omega}\right)^2\frac{1}{\cosh^2[\beta(m-g^2/\omega)/2]}.
\label{fso1a}
\end{equation}

\end{document}